\documentclass[reprint, amsmath,amssymb, aps, superscriptaddress, floatfix]{revtex4-1}

\usepackage[utf8]{inputenc}
\usepackage{subfigure}
\usepackage{xcolor}
\usepackage{bbold}
\usepackage{appendix}
\usepackage{graphicx}
\usepackage{dcolumn}
\usepackage{bm}
\usepackage{hyperref}

\newcommand{\bra}[1]{\left\langle #1 \right|}
\newcommand{\ket}[1]{\left| #1 \right\rangle}

\begin{document}

\title{High-dimensional quantum gates using full-field spatial modes of photons}

\author{Florian Brandt}
\email{florian.brandt@oeaw.ac.at}
\affiliation{Institute for Quantum Optics and Quantum Information (IQOQI), Austrian Academy of Sciences, Boltzmanngasse 3, A-1090 Vienna, Austria}
%

\author{Markus Hiekkam\"aki}
\affiliation{Photonics Laboratory, Physics Unit, Tampere University, Tampere, FI-33720, Finland}
\author{Fr\'ed\'eric Bouchard}
\affiliation{Department of Physics, University of Ottawa, 25 Templeton street, Ottawa, Ontario, K1N 6N5 Canada}

\author{Marcus Huber}
\affiliation{Institute for Quantum Optics and Quantum Information (IQOQI), Austrian Academy of Sciences, Boltzmanngasse 3, A-1090 Vienna, Austria}

\author{Robert Fickler}
\email{robert.fickler@tuni.fi}
\affiliation{Institute for Quantum Optics and Quantum Information (IQOQI), Austrian Academy of Sciences, Boltzmanngasse 3, A-1090 Vienna, Austria}
\affiliation{Photonics Laboratory, Physics Unit, Tampere University, Tampere, FI-33720, Finland}

\begin{abstract} 
Unitary transformations are the fundamental building blocks of gates and operations in quantum information processing allowing the complete manipulation of quantum systems in a coherent manner. In the case of photons, optical elements that can perform unitary transformations are readily available only for some degrees of freedom, e.g. wave plates for polarisation. However for high-dimensional states encoded in the transverse spatial modes of light, performing arbitrary unitary transformations remains a challenging task for both theoretical proposals and actual implementations. Following the idea of multi-plane light conversion, we show that it is possible to perform a broad variety of unitary operations when the number of phase modulation planes is comparable to the number of modes.  More importantly, we experimentally implement several high-dimensional quantum gates for up to 5-dimensional states encoded in the full-field mode structure of photons. In particular, we realise cyclic and quantum Fourier transformations, known as Pauli $\hat{X}$-gates and Hadamard $\hat{H}$-gates, respectively, with an average visibility of more than 90~\%. In addition, we demonstrate near-perfect ``unitarity" by means of quantum process tomography unveiling a process purity of 99~\%. Lastly, we demonstrate the benefit of the two independent spatial degrees of freedom, i.e. azimuthal and radial, and implement a two-qubit controlled-NOT quantum operation on a single photon. Thus, our demonstrations open up new paths to implement high-dimensional quantum operations, which can be applied to various tasks in quantum communication, computation and sensing schemes.

\end{abstract}

\maketitle

\section{Introduction}

In times where optical quantum information processing tasks slowly enter the realm of everyday technical applications \cite{Dowling2003,Yin2017,Kues2017,Wang2018}, the reliable and efficient control of the Hilbert-space of a quantum system becomes increasingly important. There are many optical elements known, which perform unitary transformations (acting on different degrees of freedom) on certain input modes in order to achieve a desired mode content of the output. For instance, in the polarisation space, elements such as wave plates that make use of birefringence can be used to convert the photon's polarisation states in a unitary fashion. The existence of such simple and efficient elements is one of the main reasons why this degree of freedom (DOF) has been used in a myriad of quantum experiments both in fundamental studies \cite{Giustina2015,Shalm2015,Yin2017} as well as in applications \cite{Gisin2002,Xu2019}. However, it is known that high-dimensional systems, so-called \textit{qudits}, offer access to several advantages such as an increase in channel capacity as well as an improved resistance to noise in communication protocols \cite{Cerf2001,Bechmann-Pasquinucci2000,Ecker2019} with feasible experimental effort \cite{Bavaresco2018}.

One very popular candidate for the implementation of high-dimensional information processing protocols that has gained a lot of attention in recent years is the transverse spatial DOF. 
A convenient and very popular discretization of the two-dimensional transverse space can be done by using the Laguerre-Gauss (LG) modes \cite{Erhard,Rubinsztein-Dunlop2017}, which form a complete, orthonormal basis. LG modes are characterised by a twisted helical wavefront of the form $e^{i \ell\phi}$, where $\phi$ is the azimuthal coordinate and $\ell$ corresponds to the quantized orbital angular momentum (OAM) value~\cite{Allen1992a}, which can take on values as large as 10,010~\cite{Fickler2016}. In addition, LG modes are characterised by a second quantum number associated with the radial structure, which is often labelled $p$ and only attracted notable attention recently~\cite{karimi2012radial,karimi2014radial,Karimi2014,Plick2015}.

While there are infinitely many ways to decompose the continuous spatial DOF into orthogonal modes, LG modes with their OAM value are naturally conserved in down-conversion processes, which are the workhorse of experiments in photonic quantum information processing. This makes them a natural Schmidt basis for analysing entanglement and implementing quantum communication protocols.
The family of LG modes has thus been the central matter of interest in various fields of experimental quantum information demonstrations using high-dimensional quantum states \cite{Mair2001,Krenn2014,Mirhosseini, McLaren2015,Malik2016,sit2017high,bouchard2017high}.

Here, the advances in technology to shape and detect the transverse structure of light with high precision has played a key role. In addition, the advantage of having multiple quantum states co-propagating along one optical axis, eases the implementation of more complex experimental arrangements due to intrinsic relative phase stability without the need for interferometric setups \cite{Cardano2015,Cardano2017}. 
While the accessible Hilbert space is in principle infinite-dimensional, technical hurdles such as the aperture of the optical system or the resolution of cameras and tools for the manipulation of wavefronts, limit the number of modes that can be harnessed in practical applications.

Nonetheless, experimental implementations already manage to access multiple modes with high fidelity. Indeed, for state generation, different optical techniques such as computer generated holograms \cite{Heckenberg1992,forbes2016creation} or direct modulation of the transverse phase \cite{Turnbull1996,Marrucci2006}, have been used. 
Measuring the spatial mode of single photons can also be considered a mature technology, with approaches ranging from mode sorting with the help of phase elements and free-space propagation \cite{Berkhout2010,Mirhosseini2013,Leach2002,OSullivan2012,Zhou2017,Fontaine2018a,Fontaine2017,Gu2018,Fickler2019} to mode filtering using phase- and intensity-flattening techniques along with single mode fibre coupling \cite{Mair2001,bouchard2018measuring}.

To truly harness the potential of this high-dimensional space, however, the ability to reversibly implement any transformation in the subspace spanned by a finite number of selected modes, is crucial.
And indeed, while the generation and measurement of transverse spatial modes has been investigated extensively, only a few unitary mode transformations have been realised despite their utmost importance for quantum information science.  
Early on, it was realised that cylindrical lenses arranged properly can transform a set of LG-modes into Hermite-Gauss modes and vice versa~\cite{Courtial1999}.  More recently a complex arrangement of bulk optical elements has been implemented to realise a universal gate on spin-orbit 4-dimensional states~\cite{slussarenko2009universal}, and a 4-dimensional version of the  $\hat{X}$-gate, i.e. a cyclic permutation of the input modes \cite{Schlederer2016,Babazadeh2017}. It has also been shown how to extend this to arbitrary dimensions using only linear optical elements arranged in complex interferometric setups and free-space propagation \cite{Gao2019}. Thus, together with a mode-dependent phase operation (simply performed by a Dove prism) any unitary operation on the OAM subspace could be performed, at least theoretically. 

A device which is able to perform any unitary operation on a given high-dimensional state space is often termed \textit{multiport} and has been realised for path-encoding~\cite{Schaeff2015,Wang2018} and similarly for the time-frequency domain~\cite{Kues2017,Imany2018}. For the transverse spatial degree of freedom, only one experiment so far demonstrated a single mode conversion using multi-plane phase modulation implemented by a deformable mirror~\cite{Morizur2010}. While the latter nicely demonstrates the potential of the technique, to the best of our knowledge, no experiment so far demonstrated the implementation of a flexibly programmable device that is able to perform any unitary operation between a set of input modes and a set of output modes for the full transverse spatial DOF of light.

Here, we present an experiment where we use a wavefront matching (WFM) technique~\cite{Labroille2014,Takahashi2007} to implement a fully programmable multiport for transverse spatial modes of light in a 
multi-plane light-conversion (MPLC) setting. In contrast to earlier demonstrations of sorting and multiplexing~\cite{Fontaine2018a,Labroille2014} as well as single mode conversions~\cite{Morizur2010}, we perform key quantum operations such as Pauli $\hat{X}$-gates and Hadamard $\hat{H}$-gates for multiple input and output modes taking into account both the azimuthal and radial DOF. We furthermore exploit the possibility of addressing all superposition states as inputs and outputs and perform full quantum process tomography. We find a process purity of 99~\% for cyclic transformations, which demonstrates the ``unitarity" of the performed operation, a key measure for the quality of quantum information processing tasks. Finally, we perform a quantum operation on a single photon by taking advantage of the two ``independent" transverse DOF. In particular, we perform a controlled-NOT operation using both the radial and the azimuthal DOF of single photons, thereby highlighting the benefits of the ability to control the full-field structure of single quantum systems.

\begin{figure*}[htbp]
	\begin{center}
	\includegraphics[width=0.95\linewidth]{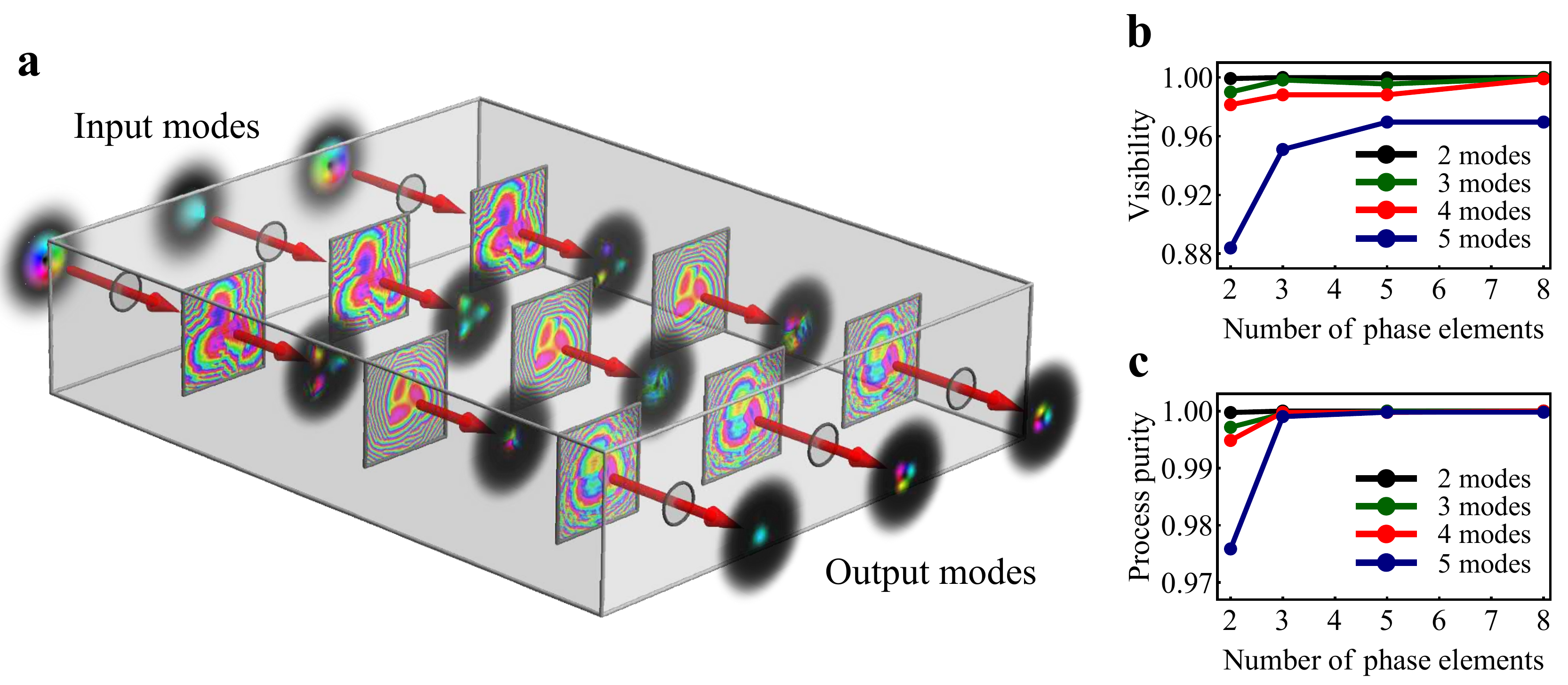}
	\caption[]{\textbf{Presentation of the multi-plane light-conversion (MPLC) technique.} \textbf{(a)} The simulated evolution of the input modes at different stage of the MPLC setup is shown for the case of three phase elements. We note that the input modes are all co-propagating and therefore experience the same three phase manipulations. The simulated \textbf{(b)} visibility and \textbf{(c)} process purity of the three-dimensional cyclic transformation is shown as a function of the number of phase elements for different numbers of OAM modes.}
	\label{fig1}
	\end{center}
\end{figure*}

\section{Unitary transformations using MPLC}

The key task is to realise an experimental implementation of a so-called multiport that is able to perform any unitary operation $\hat{U}$ on a given set of $d$-dimensional input states $\rho_{i}$, i.e. in our case a set of LG spatial modes, that converts them into a well-defined set of $d$-dimensional output states $\rho_f$:
\begin{eqnarray}
\rho_{f} = \hat{U}\rho_{i}\hat{U}^\dagger
\label{eq:Unitary}
\end{eqnarray}
The main idea behind our experimental realisation of the unitary transformation $\hat{U}$ is that we use a multi-plane phase modulation technique to build a mutliport, which shares similarities to some earlier work \cite{Morizur2010,Labroille2014}. In this process the device acts on all input modes at the same time, thus, no splitting of the modes and a separate phase-stable modulation is required. Instead of a stochastic optimization algorithm used earlier~\cite{Morizur2010}, we adapt the technique of wavefront matching (WFM), which we outline below. 

\subsection{Wavefront matching (WFM)}
The WFM technique is known from waveguide design \cite{Takahashi2007} and has recently been used to perform LG mode sorting using multiple phase modulations~\cite{Fontaine2018a}. During the WFM optimisation, all $d$ input modes $f_r$ are propagated forward through an optical system containing $n$ phase elements $\Phi_t$ with some free space propagation in between. At each modulation plane, $t=1,...,n$, the complex amplitudes of all modes $f_r(x,y,t)$ are recorded. Note that we do not perform a full Fourier transform between the phase modulations but use a split-step technique to propagate through the system. Subsequently, all final output or target modes $b_s$ are propagated backwards, first to the last phase modulation plane ($t=n$) to obtain $b_s(x,y,n)$. Now all input-output mode pairs, i.e. $f_r$ and $b_s$ where $r=s$, are ``compared" to find the best single phase modulation that matches all wavefronts at the same time. For this, a field overlap between each input-output mode pair is calculated according to $o_{rst}(x,y)=\overline{b_s(x,y,t)} f_r(x,y,t) e^{i\Phi_t(x,y)}$ including a transverse phase modulation $\Phi_t(x,y)$, which is set to zero in the first iteration round but will be updated during the WFM process. Subsequently, the required phase patterns for the plane $t$ can be obtained through: 
\begin{eqnarray}
\Delta\Phi_t(x,y) = - \arg \left( \sum_{r=s} o_{rst}(x,y) e^{-i\phi_{rst}}  \right),
\label{eq:PhaseMod}
\end{eqnarray}
where $\phi_{rst}$ is the average phase of the calculated overlap for each mode pair. The resulting phase modulation for the last plane $\Phi_n(x,y)$ is then imprinted on each backwards propagating mode $b_s$, which is then propagated to the $(n-1)$th modulation plane. Note that due to the free-space propagation between two phase modulations, the amplitude is slowly adjusted to match the input and output modes. At plane $(n-1)$, the overlaps between all input-output mode pairs are calculated again and the required phase change $\Delta\Phi_{n-1}(x,y)$ is obtained through the formula given in Eq.~(\ref{eq:PhaseMod}) and imprinted on all backwards propagating modes. Then, this procedure of propagating, comparing and phase matching is repeated until the very first plane. During the subsequent propagation and modulation, the wavefronts of the backwards propagating modes approach the input modes with respect to their phase and amplitude until they are perfectly matching in case of a very large number of phase elements $n$. If the number of phase modulations is limited, one iteration often only leads to a weak resemblance of the input-output mode pairs. However, the whole procedure can be repeated as many times as necessary until a certain fidelity is reached. We found that for all our calculations, we only require a maximum of 50 iterations after which the mode overlaps between input and output modes do not improve anymore while numerical errors slowly deteriorate the result. As the input and output modes can be chosen freely, we now established a procedure to implement any unitary operation between these modes, hence realising a multiport using only a limited number of phase elements placed along the optical axis. 

Before investigating how many phase elements are required to obtain reasonably good results, we briefly discuss a few important unitary operations, which play a key role as important quantum gates in the field of quantum computation and quantum communication.

\subsection{High-dimensional quantum gates}
Some of the most important unitary operations $\hat{U}$ in terms of high-dimensional quantum gates are the Pauli $\hat{X}$-gates, Hadamard $\hat{H}$-gates as well as controlled Pauli c$\hat{X}$-gates. The first two operations are single qudit operations and corresponds to a cyclic transformation and a quantum Fourier-transform, respectively. The third gate is a two-qudit operation, where one high-dimensional quantum systems controls the cyclic-operation on another high-dimensional quantum state. Note that for the latter, the two-qubit version is also known as a controlled NOT operation.

\medskip
\textbf{$\hat{X}$-gate}:\medskip

The effect of a high-dimensional $\hat{X}$-gate (i.e cyclic transformation) on a certain $d$-dimensional quantum state can be mathematically expressed as
\begin{eqnarray}
\hat{X}^m\ket{l}= \ket{(l+m)_{\text{mod}(d)}}
\label{eq:Xgate}
\end{eqnarray}
which simply corresponds to a cyclic operation where each mode gets transformed to its $m$-th nearest neighbour-mode, modulo the number of modes $d$. Interestingly, it was shown only recently and only for the OAM degree of freedom that this operation can be implemented for arbitrary dimensions using a complex arrangement of linear optical elements and free-space propagation \cite{Gao2019}. For full-field modes, i.e. including both the azimuthal and the radial modes, no implementation of cyclic transformations is known. Examples of matrix representations of these gates for dimension three, i.e. qutrits, can be found in Appendix A.

\medskip
\textbf{Hadamard-gate}:\medskip

The high-dimensional quantum Fourier-transform operation is also known as a high-dimensional Hadamard $\hat{H}$-gate due to its relation to the well-known Hadamard gate for qubits. Its mathematical representation for prime dimensions (larger than two) can be formulated as
\begin{eqnarray}
\hat{H}_m\ket{\Psi_l} = \frac{1}{\sqrt{d}} \sum_{k=0}^{d-1}\omega^{(lk+(m-1)k^2)}_d\ket{k},
\label{eq:Hadamard}
\end{eqnarray}
where $\omega_d=e^{i 2\pi /d}$ and $m=1,...,d$ and $\hat{H_0}$ is the identity. This operation transforms any given input state of a basis into a coherent superposition of all the modes of that basis with different well-defined phases. The different configurations of the Hadamard gate $\hat{H}_m$ can be seen as a simple change between mutually unbiased bases (MUB). While the switching among all possible MUBs is important for various quantum tasks, e.g. for quantum state tomography and cryptography~\cite{Scarani2009}, there is one basis that is intuitively simple to understand for LG modes with $p=0$ and OAM values symmetrically distributed around $l=0$, because it corresponds to a transformation into angular coordinates. Again, we note that there is no bulk optics realisations known for such operations. For the four different Hadamard transformation matrices for qutrits, we refer to the Appendix A.

\medskip
\textbf{Controlled $\hat{X}$-gate}:\medskip

The final quantum operation we discuss here and implement in our experiment below, is a gate requiring two high-dimensional quantum states, where one is used to control the other, i.e. a controlled $\hat{X}$-gate or c$\hat{X}$-gate. Its mathematical representation can be formulated as follows:
\begin{align}
\text{c}\hat{X}\left(\ket{p} \ket{l} \right) = \left(\hat{I}\otimes \hat{X}^p \right) \ket{p}\ket{l} = \ket{p}\ket{(l+p)_{\text{mod}(d)}}
\label{eq:CNOT}
\end{align}
The commonly known qubit version is the so-called controlled NOT-gate (cNOT), where one qubit acts as a control qubit for a second target qubit, on which a NOT-operation is performed if and only if the control qubit is $\ket{1}$ and leaves it unchanged if it is $\ket{0}$. Generalised to arbitrary dimensions for both, the control as well as target qudit, the operation acts as a cyclic operation on the target qudit, where the state is shifted by the value of the control qudit. In our case, one qudit, e.g. the control qudit, can be realised by the radial DOF of light, while the target qudit corresponds to the OAM value. Thus, if the input photon has a radial structure of $p$ the OAM value $l$ gets shifted by $p$ modulo the dimension of $l$. In the appendix A we give the transformation matrix of a two-dimensional cNOT gate as an example.

\begin{figure*}[htbp]
	\begin{center}
	\includegraphics[width=0.95\linewidth]{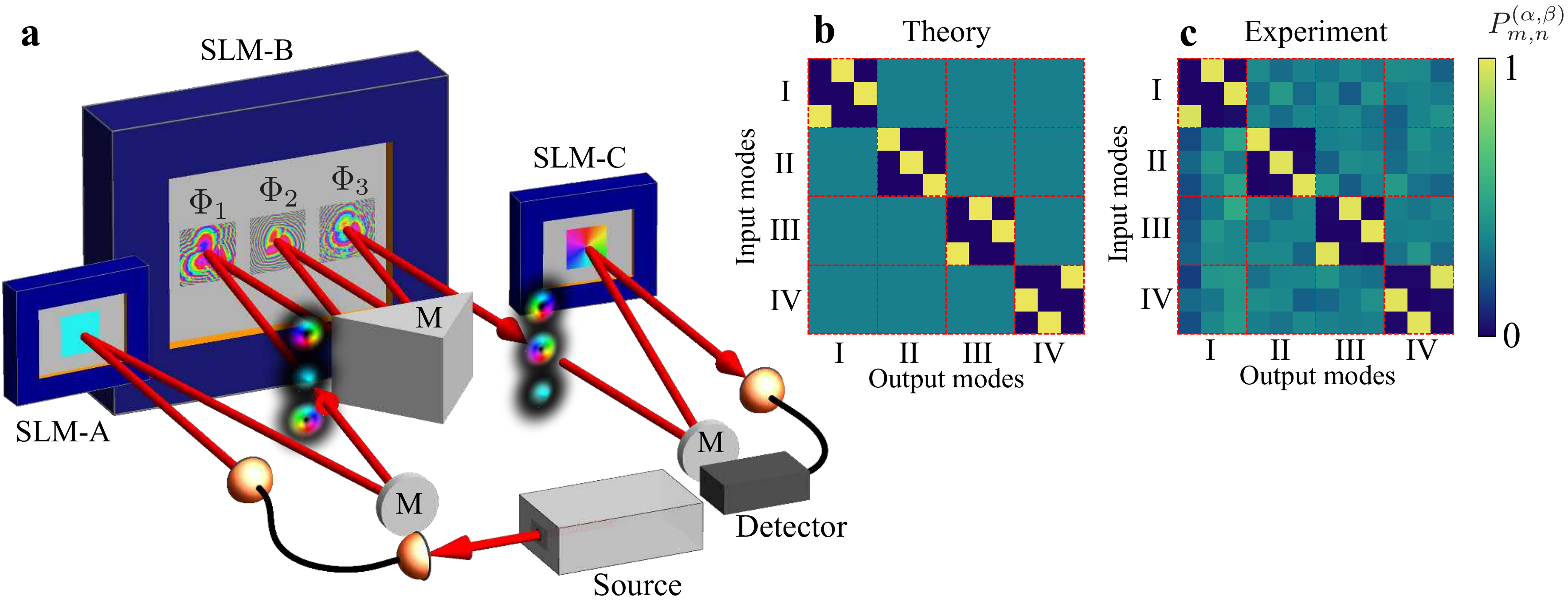}
	\caption[]{\textbf{Experimental implementation of the three-dimensional cyclic transformation.} \textbf{(a)} Simplified experimental setup showing the implementation of the three-dimensional $\hat{X}$-gate using 3 phase elements. Photons from the source are spatially filtered to the fundamental Gaussian mode using a single mode fibre (black wire). The beam is then made incident on SLM-A, where both the phase and amplitude is manipulated in order to generate the appropriate LG mode. These modes are then fed into the MPLC system consisting of SLM-B, displaying three phase patterns, and a mirror (M). The beam bounces off between the SLM and the mirror three times before it exits the system and gets detected using SLM-C with the intensity flattening technique introduced only recently in \cite{Bouchard2018}. \textbf{(b)} Theoretical prediction of a tomographically complete set of measurements for the theoretical three-dimensional $\hat{X}$-gate. The input and output states are chosen from the 3-dimensional MUBs labelled here by I, II, III and IV. \textbf{(c)} Experimental correlation matrix for different MUBs obtained from the MPLC system described in \textbf{a}.}
	\label{fig2}
	\end{center}
\end{figure*}

\subsection{Implementation using MPLC}

We now turn to the investigation of how many planes are required to realise the above discussed $d$-dimensional quantum gates using the multiplane phase conversion. First, we calculate the phase modulations using 2,3,5 and 8 phase elements for up to 5 OAM modes for the $\hat{X^1}$-gates and $\hat{H}_{d}$-gates using the WFM method. As an example of one simulated unitary operation, we show the scheme of a qutrit $\hat{X}^1$-gate for the LG modes with $p=0$ and $l=0,\pm1$ using three planes of phase modulation in Fig.~\ref{fig1}-\textbf{a}. Similar to~\cite{Morizur2010,Labroille2014}, we find that if the number of modes exceed the number of planes, the quality of the transformation decreases significantly. As a measure of the quality of the transformation, we use the visibility $V$, which we obtain from the cross talk between the modes according to: 
\begin{eqnarray}
V= \sum_i C_{ii} / \sum_{ij} C_{ij}, 
\label{eq:Vis}
\end{eqnarray}
where $C_{ij}$ corresponds to the probability entries in the diagonal cross-talk matrix and $C_{ii}$ signifies the probability of the input mode being transformed into the desired output mode.
Although it is usually a good rule-of-thumb that one needs twice as many planes as modes, we already obtain a very low cross-talk, i.e. a visibility $V$ in excess of 96~\%, when the number of modes is equal to the number of planes and at least three planes are utilised, see Fig.~\ref{fig1}-\textbf{b}. Since simple cross-talk matrices cannot directly reveal information about the unitarity of the mode transformation, we also perform a full high-dimensional quantum process tomography~\cite{Bouchard2018} on the simulated 3-dimensional $\hat{X}^1$-gate. 

Quantum process tomography is based on a set of informationally complete measurements for a given set of input states, leading to a complete characterisation of the corresponding quantum channel. The channel can then be represented by a completely positive map ${\cal E}$, where $\rho_f={\cal E}\left(\rho_i\right)$. Using a fixed set of operators, ${\cal E}$ can be expressed as
\begin{eqnarray}
{\cal E}\left(\rho \right) = \sum_{i,j} \chi_{i,j} \hat{\sigma}_i \, \rho \, \hat{\sigma}_j^\dagger,
\end{eqnarray}
where $\chi$ is a $d^2 \times d^2$ dimensional matrix known as the process matrix, and $\hat{\sigma}$ are the Gell-Mann matrices, also known as Pauli matrices for $d=2$. Hence, this representation of quantum processes can describe more general processes, e.g. decoherence of the input state, than for the case of unitary transformations ($\mathrm{rank}\left(\chi \right)=1$). The unitarity of the transformation can be assessed by taking advantage of the Choi-Jamiolokowski isomorphism, which implies that ${\cal E}$ can be represented by an operator $\rho_{\cal E}$, known as the Choi matrix, given by
\begin{eqnarray}
\rho_{\cal E}=\left( \hat{I} \otimes {\cal E} \right) \ket{\Psi} \bra{\Psi},
\end{eqnarray}
where $\ket{\Psi}$ is a $d$-dimensional maximally entangled state. Thus, the \emph{process purity}, $\mathrm{Tr}\left[\rho_{\cal E}^2 \right]$, is a measure of the extent to which the purity of the input states are maintained throughout the quantum process~\cite{gilchrist2005distance}.  

In order to perform process tomography, we propagate all states of all MUBs through the device, compare the simulated outcome with the targeted modes to obtain a correlation matrix, from which one can deduce key figures of merit, such as the process purity, which is of key importance for all quantum information processing tasks. 
 
Having shown that only a few phase planes are enough to enable a broad range of quantum gates in simulations, we now implement a simple experimental realization of a quantum gate.

\section{Experimental results}

\subsection{$\hat{X}$-gate using 3 phase modulations and OAM-qutrits}
\begin{figure}[htbp]
	\begin{center}
	\includegraphics[width=1\linewidth]{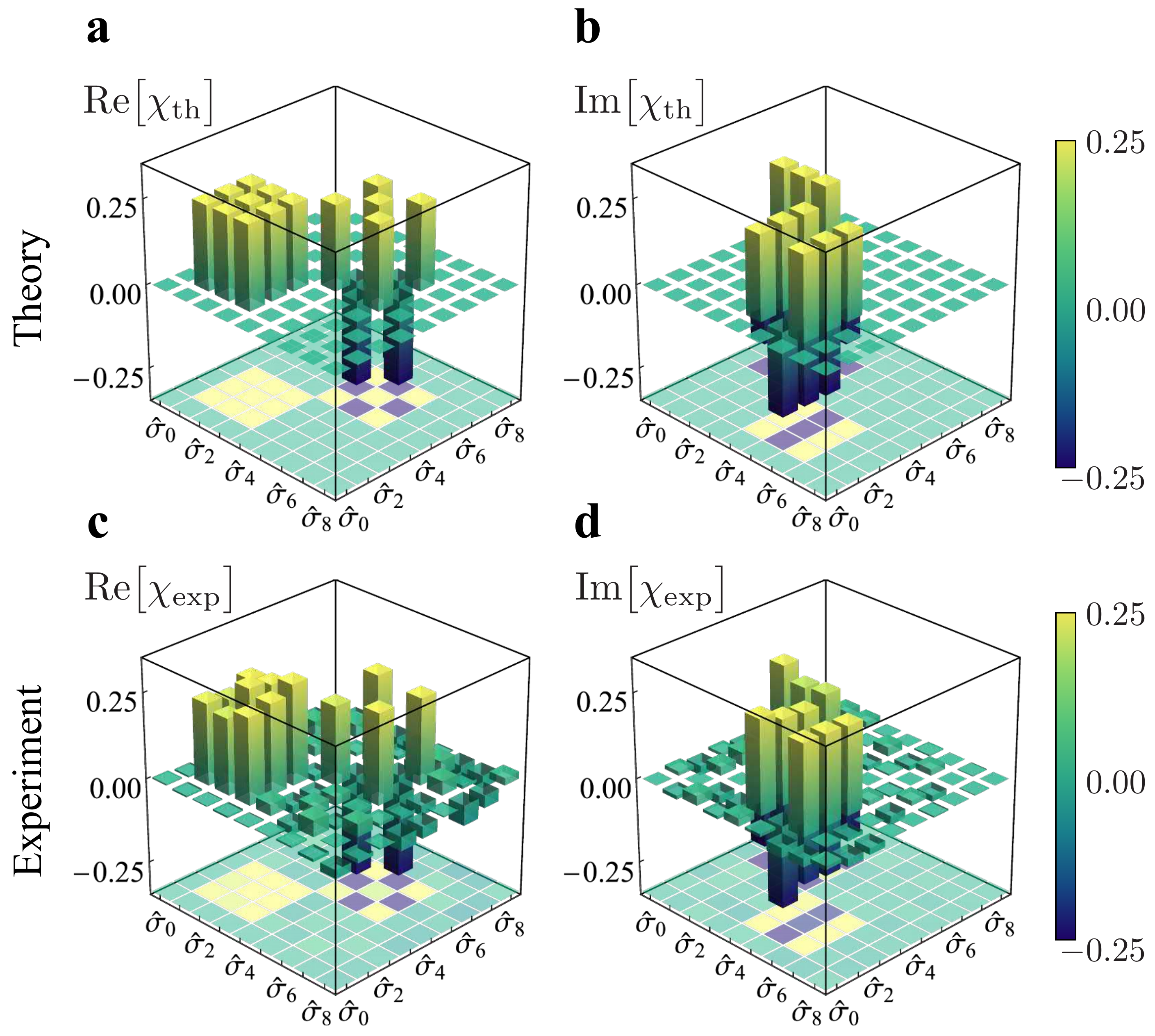}
	\caption[]{\textbf{Quantum process tomography of the three-dimensional cyclic transformation.} The real \textbf{(a)} and imaginary \textbf{(b)} part of the theoretical process matrix, $\chi_\mathrm{th}$, for the three-dimensional $\hat{X}$-gate are shown in the Gell-Mann matrix basis. The real \textbf{(c)} and imaginary \textbf{(d)} part of the experimentally reconstructed process matrix, $\chi_\mathrm{exp}$, for the experimental implementation of the three-dimensional $\hat{X}$-gate using 3 phase elements and OAM modes $\ell=-1,0,1$ are shown in the Gell-Mann matrix basis.}
	\label{fig3}
	\end{center}
\end{figure}

As a first experimental test and benchmark of a multiport realised by MPLC, we implement the 3-dimensional $\hat{X}^1$-gate for LG modes ${\{\ket{l=\text{-}1},\ket{l=0},\ket{l=1}\}}$ and ${p=0}$ by using only 3 modulation planes. We start with a low number of modes and phase modulations to investigate a best-possible implementation in terms of efficiency, resolution and modulation ability. We further use only LG modes with no radial structure to minimise errors introduced by the detection system~\cite{bouchard2018measuring}. A sketch of the experimental setup can be seen in Fig.~\ref{fig2}-\textbf{a}. At first we imprint the required LG mode structures on a spatially cleaned 808~nm laser beam with a strongly enlarged Gaussian mode and carve out the required LG mode by modulating the phase and amplitude of the light using a single spatial light modulator (SLM-A)~\cite{Bolduc2013b}. We then implement the multiport by another SLM (SLM-B) in combination with a mirror opposing it. The laser is sent through this arrangement such that it bounces off the SLM three times, which enables three separate phase modulations. Each of the three phase modulations covers a large area of $630 \times 630$~pixels on the SLM to reduce errors introduced by the finite resolution of the holograms. We further reduce errors introduced by slight (pixel-sized) misalignments by using a beam waist of around 1~mm that is much larger then the pixel pitch (8~$\mu$m). The propagation distance between each phase modulation, i.e. each reflection on the SLM, is set to 800~mm, thus ensuring enough propagation for phase-induced amplitude modulation and a proper functioning of the mode conversion. The utilised phase modulations are the ones presented in Fig.~\ref{fig1}. We note that in our experiment, we additionally imprint a diffraction grating and use only the first diffraction order thereby filtering out the unmodulated light remaining in the zeroth order. We additionally decrease the beam waist of the modes during the transformation by a factor of 2 (implemented by the WFM code), to have a smaller beam impinging on the third spatial light modulator (SLM-C) in order to improve the measurement, which uses phase- and intensity- flattening of the beam to ensure, although lossy, near-perfect spatial mode projections~\cite{bouchard2018measuring}. With this configuration, we achieve a three-dimensional $X^1$-gate with a visibility $V$ of $(98.4\pm 0.7)~\%$ between all 144 input-output mode combinations of all four MUBs (see full cross-talk matrix in Fig.~\ref{fig2}-\textbf{c}). This result is close to the one expected from simulation of 99.6~\% (Fig.~\ref{fig2}-\textbf{b}) and shows the near perfect functioning of the experimental implementation of our multiport. Furthermore, we perform high-dimensional quantum process tomography of the experimental three-dimensional $\hat{X^1}$-gate for which the theoretical process matrix can be seen in Fig.~\ref{fig3}-\textbf{a}-\textbf{b}. The experimentally reconstructed process matrix is shown in Fig.~\ref{fig3}-\textbf{c}-\textbf{d}. We find a process purity of 99.3~\%, which is in perfect agreement with the simulated prediction. Moreover, the process purity nicely shows that the transformation is fully coherent and as such a powerful tool in quantum information schemes.

\subsection{High-dimensional gates for OAM modes}

As we have shown earlier, a larger number of phase modulations allows unitary operations on a larger mode set and as such a larger dimension of the quantum state. To test the limitations in terms of simplicity of the experimental implementation and maximum number of phase modulations, we now realise a setup with a multiport consisting of 8 phase modulations, as well as the generation and detection on the same SLM screen, i.e. a setup where the beam bounces off the SLM 10 times in total. In contrast to the experimental setup shown in Fig.~\ref{fig2}-\textbf{a}, we now implement five reflections on the upper half of the SLM, each phase modulation being $160 \times 160$~pixels in size. Here, we use a beam of around 0.5~mm waist and we propagate it only 100~mm between each phase modulation (50~mm between mirror and SLM), to keep the whole setup more compact. After the fifth reflection, we insert another slightly tilted mirror into the setup, which sends the light back onto the lower part of the SLM, where the beam bounces off the SLM another five times before it can leave the MPLC arrangement. Since we have to redirect the light from the upper to the lower half of the SLM, we compensate this vertical redirection of the beam by displaying additional vertical grating structures on the five holograms displayed on the lower half of the SLM. For the generation (done by the first hologram) and the detection of the modes (last hologram) we again use amplitude and phase modulation~\cite{Bolduc2013b} as well as phase- and intensity- flattening~\cite{bouchard2018measuring}. As there are now 10 relatively small holograms in total, a misalignment of only one pixel for each hologram leads to a significant reduction of the quality of the mode transformation. Moreover, when a larger number of modes is involved, the phase modulations calculated by the WFM tend to get more complicated, which makes the alignment even harder. Due to this fact, we make use of an automated alignment procedure based on a genetic algorithm~\cite{haupt2004}, where each member of the population defines the position of all holograms and the feedback signal is given by the visibility $V$ of the cross-talk measurements of the transformation as defined in Eq.~(\ref{eq:Vis}). With this programmable and fully automated multiport, we are now able to test various transformations and single-qudit gates for different modes and dimensions.

In a first set of measurements, we only use LG modes with radial index $p=0$.
For a three-dimensional state space spanned by the three lowest OAM modes, i.e. the set \{${\ket{l=\text{-}1},\ket{l=0},\ket{l=1}}$\}, we experimentally obtain an average visibility of $(92.7\pm 3.8)~\%$ for all three possible $\hat{X}$-gates in the computational basis (MUB I). We further measure an average visibility of $(92.0\pm3.0)~\%$ for cyclic transformations between different MUBs, which corresponds to the unitary transformation $\hat{U}=\hat{X}\hat{H}$, i.e. a combined Pauli $\hat{X}$-gate and Hadamard $\hat{H}$-gate  operation. The exact results of all recorded correlation matrices for this as well as for all following measurements can be found in the Appendix B.

\begin{figure}[htb]
	\begin{center}
	\includegraphics[width=1\linewidth]{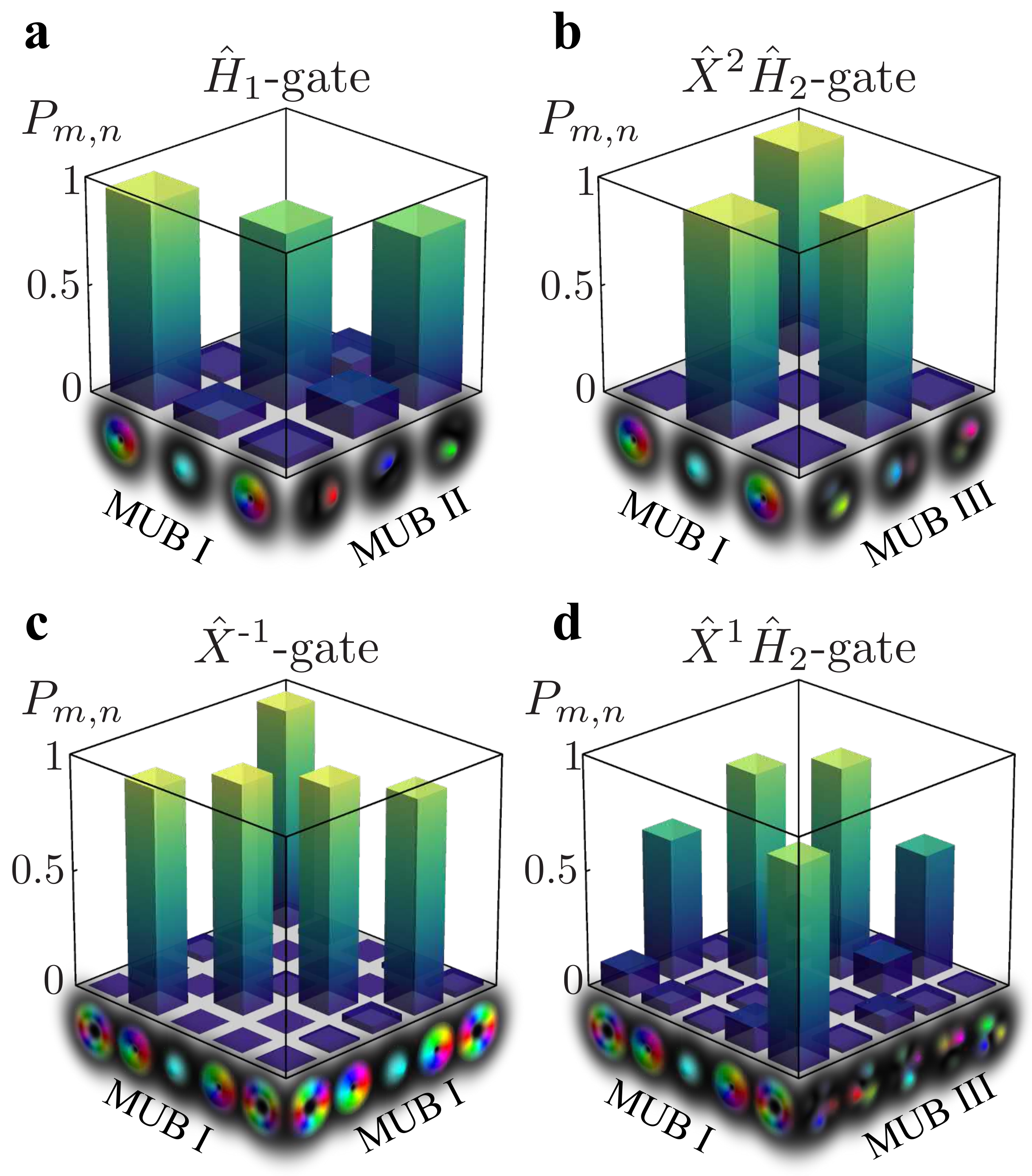}
	\caption[]{\textbf{Experimental results for the OAM transformations using a single SLM for generation, manipulation (8 phase planes) and detection.} Crosstalk matrices are shown for a three-dimensional Hadamard-gate ($\hat{H}_1$) converting input beams from the computational basis to another mutually unbiased basis (here the angle basis) \textbf{(a)} and a Pauli $\hat{X}^1$-gate in dimension five \textbf{(c)}. We further realized  combined $\hat{X}\hat{H}$ operations performing a cyclic operation on a set of input modes that is transformed to another MUB in dimension three  ($\hat{X}^2\hat{H}_2$-gate) \textbf{(b)} and dimension five ($\hat{X}^1\hat{H}_2$-gate) \textbf{(d)}. Note that the quality for measurements of complex superposition modes in large Hilbert spaces as in \textbf{(d)} decreases due to limited resolution of the holograms for the single-SLM implementation.}
	\label{fig4}
	\end{center}
\end{figure}

We then increased the dimension to $d=5$ by including second order OAM modes, i.e $\ket{l=\text{-}2}$ and $\ket{l=2}$. We obtained an average visibility over all $\hat{X}$-gates in the computational basis of $(91.9\pm3.2)~\%$. However, for the modes of the other MUBs, the obtained visibility of the cross-talk matrix does not reach the theoretical value but is found to be only 76.5~\% (see fig.\ref{fig4}d). We relate this decrease in transformation quality to the finite resolution of our SLM, i.e. to be only of technical nature. The reason being that the complexity of the spatial structure of higher-order superposition modes (the states of the other MUBs), i.e. the field gradients, gets comparable in size to the pixels of the SLM.  Examples of several recorded correlation matrices for different OAM transformations can be seen in figure \ref{fig4}.

\subsection{High-dimensional gates for radial modes}

\begin{figure}[htb]
	\begin{center}
	\includegraphics[width=1\linewidth]{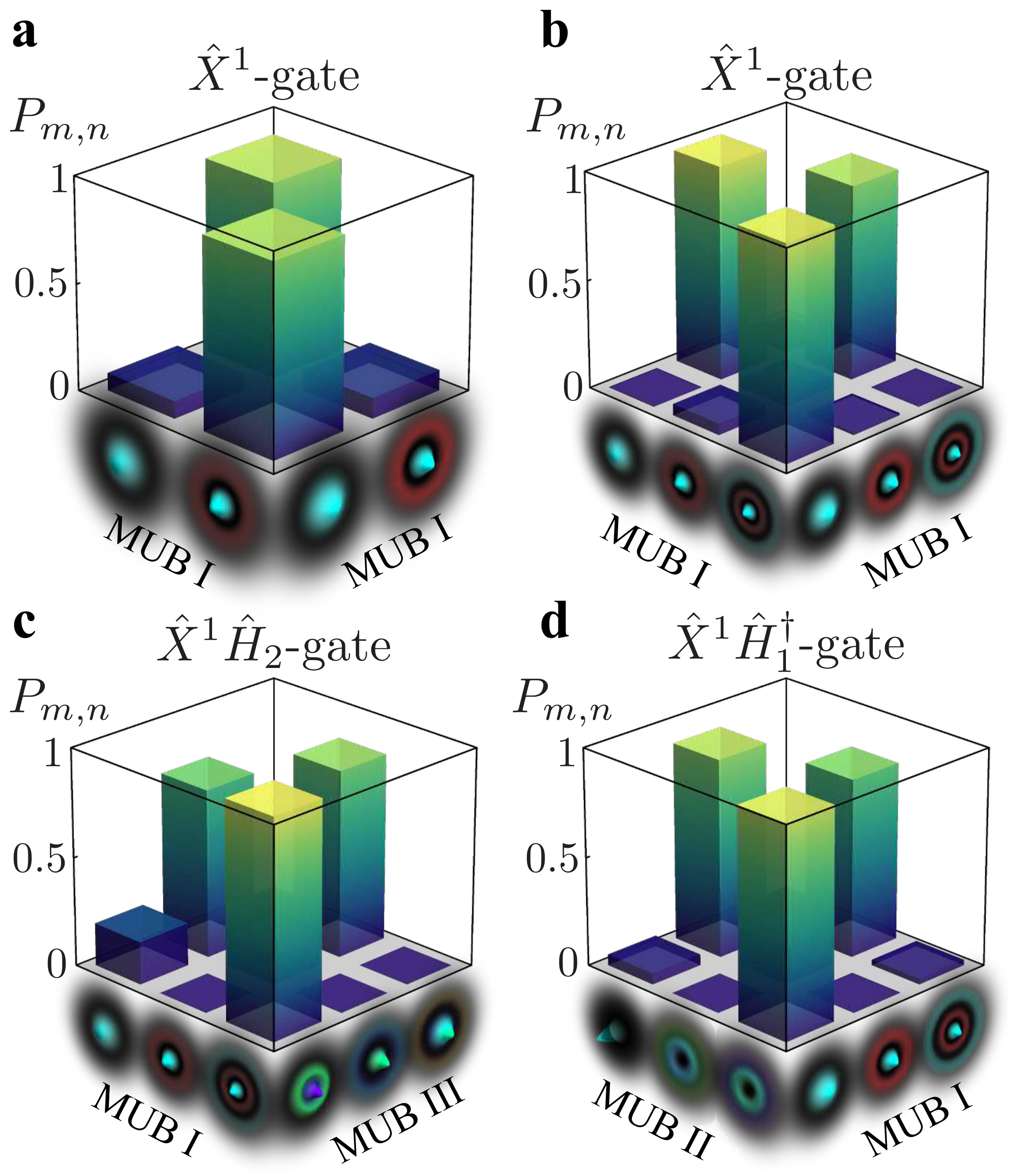}
	\caption[]{\textbf{Experimental results of different p-mode transformations, again using only one SLM for generation, manipulation, and detection.} A cyclic transformation $\hat{X}^1$ on $p$-only modes in the computational basis  is shown for dimension two in \textbf{(a)} and dimension three in \textbf{(b)}. In addition, we implement also combined $\hat{X}\hat{H}$-gates in dimension three, in particular a $\hat{X}^1\hat{H}_2$-gate shown in \textbf{(c)} and a $\hat{X}^1\hat{H}_1^\dagger$-gate shown in \textbf{(d)}. Note that measuring p-modes properly in different mutually unbiased bases has only become possible recently~\cite{Bouchard2018}.}
	\label{fig5}
	\end{center}
\end{figure}

While the OAM transformations performed above can be (at least in theory) realised using bulk optical elements~\cite{Gao2019}, we now turn to mode transformations of radial modes, i.e. $p$-modes, a task for which no other implementation is known so far. In particular, this experimental demonstration is now possible due to the recently developed measurement technique, known as intensity-flattening~\cite{bouchard2018measuring}, which enables the detection of $p$-modes in all MUBs and, thus, made it possible to perfectly measure the full field structure of LG-beams with only a minor experimental trade-off, i.e. additional loss. We performed $\hat{X}$-gates in the $p$-only space of LG-beams ($l=0$) for qubits and qutrits using the set \{$\ket{p=0},\ket{p=1}$\} and \{$\ket{p=0},\ket{p=1},\ket{p=2}$\}, respectively. Additionally, we performed $\hat{H}$-gates and combined $\hat{X}\hat{H}$-gates for the same mode set (see Fig.~\ref{fig5} for examples). The average visibility obtained over all transformations performed is $(92.4\pm3.4)~\%$, which nicely shows that our experimentally implemented multiport is not only able to perform OAM transformations but also operations on full-field modes including both the azimuthal and radial DOF, an advantage that we harness in the following section.

\subsection{Single photon controlled-$\hat{X}$ gate}
As a final test, we perform the controlled-$\hat{X}$ operation, i.e. the c$\hat{X}$-gate, introduced earlier. In contrast to the usual implementation using two quantum systems, we use two spatial degrees of freedom of a single quantum system. In order to perform this task on an actual quantum system, we exchange the laser with a heralded single photon source. The single photons are generated by a photon pair source realised by a type II parametric down conversion process (ppKTP nonlinear crystal) pumped by a 405~nm laser source. One of the two photons acts as a trigger signal to herald the existence of the single photon on which the quantum operation is performed.

\begin{figure}[htb]
	\begin{center}
	\includegraphics[width=0.9\linewidth]{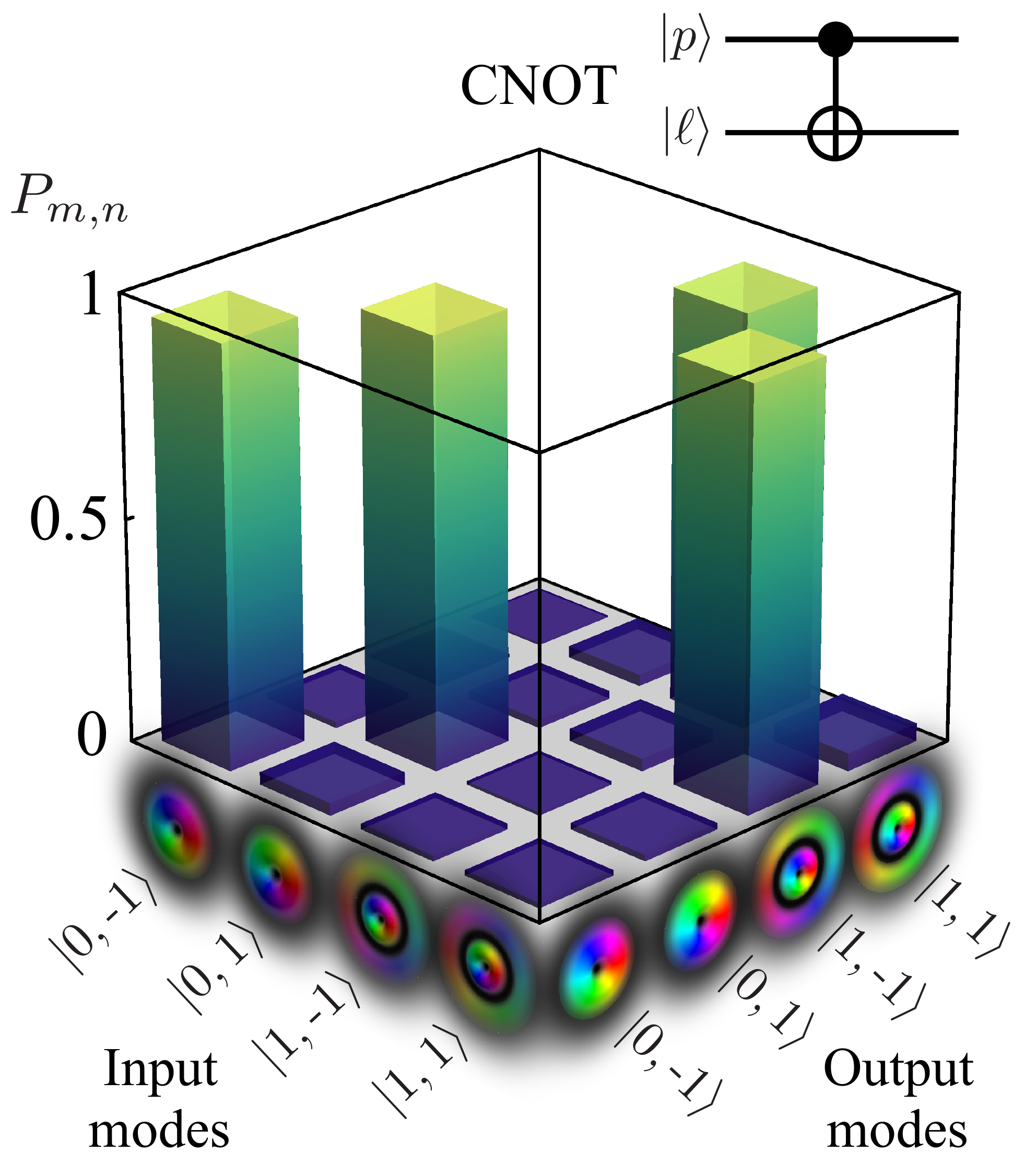}
	\caption[]{\textbf{A single-photon CNOT gate using the radial modes $\ket{p}$ as a control qubit for the manipulation of the azimuthal modes $\ket{\ell}$.} Experimental crosstalk matrix with input and output modes labelled by the states $\ket{p,\ell}$. A visibility of 94.6~\% is achieved for the single photon CNOT gate with a MPLC system consisting of a generation SLM, an SLM used for the manipulation consisting of 3 phase planes and one SLM for detection SLM. The quantum circuit diagram of a CNOT gate as used frequently in quantum information science is also shown in the top right corner.}
	\label{fig6}
	\end{center}
\end{figure}

In our demonstration, we perform the two dimensional version of this gate, which is one of the most important quantum operations, i.e. the controlled NOT gate. The control qubit is encoded in the radial structure of the photon \{$\ket{p=0},\ket{p=1}$\}, while the target qubit corresponds to the OAM quantum number \{${\ket{l=\text{-}1},\ket{l=1}}$\}.  To put it into simple terms: The cNOT operation performs a cyclic transformation on the OAM value, depending on the radial quantum number. For efficiency reasons, we realise this single photon quantum gate with the first configuration of the multiport, i.e. using only 3 phase modulations on the manipulation SLM (SLM-B) and performing the generation and detection on two additional, separate SLMs (SLM-A and SLM-C). According to simulations, three phase modulation planes already allow a visibility of 96~\% and, more importantly, increase the efficiency approximately by a factor of 5 due to a reduced number of reflections on the SLM. The resulting correlation matrix is shown in Fig.~\ref{fig6}, which corresponds to a visibility of $(94.7\pm1.4)~\%$, which is very close to the theoretical prediction.

\section{Conclusion and Outlook}

In conclusion, we have proposed and experimentally demonstrated an easy to implement, versatile scheme that can perform in principle arbitrary unitary transformations on spatial modes of light. Our method has the advantage of being simple and straightforward to implement, therefore making it a novel experimental tool for future experiments especially in the field of quantum information processing. In order to demonstrate the versatility of our technique, we demonstrated applications of different quantum gates on 3- and 5-dimensional LG modes, taking into account both the radial as well as the azimuthal quantum number. Besides the conventional high-dimensional versions of the Pauli $\hat{X}$-gate and the Hadamard $\hat{H}$-gate, we also implemented a single photon cNOT-gate, harnessing one transverse coordinate to control the other one. Moreover, we performed quantum process tomography on one particular transformation, unveiling the ``unitarity" of the operation. 

A central challenge that will have to be addressed in the future is the limited single photon efficiency, due to losses incurred by every reflection upon the SLM. For fixed quantum information protocols (e.g. QKD, where only few bases, fixed by the protocol are required) or fundamental tests this downside can be overcome by physically designing phase plates corresponding to the individual phase modulations. This would lead to high-fidelity and low-loss arbitrary unitary transformations, but would sacrifice the flexibility to program any transformation and requires a pre-designed piece of equipment for every basis. A solution which is still flexible and can be used as a re-programmable multiport, can be the implementation using deformable mirrors. They usually have very high reflection efficiencies, however, can become very costly when high spatial resolutions of the modulation is required. Nevertheless, our implementation adds another important tool for high-dimensional quantum information experiments using spatial modes of light and shows the potential spatial modes offer, especially if both the radial and azimuthal degree of freedom are taken into account.

\begin{acknowledgments}

We thank A. Zeilinger and E. Karimi for many fruitful discussions.  FBr, RF and MHu acknowledge funding from the Austrian Science Fund (FWF) through the START project Y879-N27 and the joint Czech-Austrian project MultiQUEST (I 3053-N27 and GF17-33780L).
FBo acknowledges the support of the Vanier Canada Graduate Scholarships Program and the Natural Sciences and Engineering Research Council of Canada (NSERC) Canada Graduate Scholarships program. RF and MHi acknowledge the support of the Academy of Finland through the Competitive Funding to Strengthen University Research Profiles (decision 301820) and the Photonics Research and Innovation Flagship (PREIN - decision 320165).
\end{acknowledgments}

\begin{appendix}\label{Appendix}

\section{Exemplary matrix representations of quantum gates}
\subsection{$\hat{X}$-gate}
As an example of the unitary operations in matrix representation we show all three $\hat{X}^m$-gate configurations for a (three-dimensional) qutrit:
\begin{equation*} \label{eq:XgateMatrix}
\begin{split}
\hat{X}^0 &=  \begin{bmatrix} 1 & 0 & 0  \\ 0 & 1 & 0  \\ 0 & 0 & 1   \end{bmatrix}, \quad 
\hat{X}^1 =  \begin{bmatrix} 0 & 1 & 0  \\ 0 & 0 &1  \\ 1 & 0 & 0   \end{bmatrix}, \quad 
\hat{X}^2 =  \begin{bmatrix} 0 & 0 & 1 \\ 1 & 0 &0  \\ 0 & 1 & 0 \end{bmatrix},\\
\end{split}
\end{equation*}
where the first transformation $\hat{X}^0$ is the trivial case, i.e. the identity. Note, that these matrices directly correspond to unitary mode transformations for which we optimise our multiport with respect to a given set of input modes.

\subsection{$\hat{H}$-gate}
Here again, as an example we provide the matrix representation of the $(d+1)$ $\hat{H}_n$ transformations for qutrits:
\begin{equation*} \label{eq:XgateMatrix}
\begin{split}
\hat{H}_0 &=  \begin{bmatrix} 1 & 0 & 0  \\ 0 & 1 & 0  \\ 0 & 0 & 1   \end{bmatrix}, \quad 
\hspace{0.6cm}\hat{H}_1 =  \frac{1}{\sqrt{3}}\begin{bmatrix} 1 & 1 & 1  \\ 1 & \omega_3 & \omega_3^2  \\ 1 & \omega_3^2 & \omega_3   \end{bmatrix}, \\
\hat{H}_2 &=  \frac{1}{\sqrt{3}}\begin{bmatrix} 1 & 1 & 1  \\ \omega_3 & \omega_3^2 & 1  \\ \omega_3 & 1 & \omega_3^2   \end{bmatrix}, \quad
\hat{H}_3 =  \frac{1}{\sqrt{3}}\begin{bmatrix} 1 & 1 & 1  \\ \omega_3^2 & 1 & \omega_3  \\ \omega_3^2 & \omega_3 & 1   \end{bmatrix}, \\
\end{split}
\end{equation*}
with $\omega_3=e^{i 2\pi /3}$ and the first transformation $\hat{H}_0$ again being the identity. Note that the Hadamard transformation $\hat{H}_1$ corresponds to a transformation to the angle basis (ANG) if the input modes are LG modes with no radial structure $p=0$ and OAM values symmetrically distributed around $l=0$.  

\subsection{c$\hat{X}$-gate}
Using both, the radial and azimuthal DOF, it is possible to realise a high-dimensional controlled operation on a single quantum system. As discussed in the main text, we implement the simplest version, i.e. a controlled NOT operation, where the OAM value is inverted depending on the radial structure. In our experiment we use the following set of mode combinations $\{\ket{0}\ket{-1},\ket{0}\ket{1},\ket{1}\ket{-1},\ket{1}\ket{1}\}$, where the position of the ket-vectors label the radial and azimuthal mode indices, i.e. $\ket{p}\ket{l}$, which leads to a transformation matrix of the form
\begin{equation*} \label{eq:CNOT}
\operatorname{CNOT} =  \begin{bmatrix} 1 & 0 & 0 & 0 \\ 0 & 1 &0 & 0 \\ 0 & 0 & 0 & 1 \\  0 & 0 & 1 & 0 \end{bmatrix}
\end{equation*}
We note that the implemented CNOT-gate corresponds to the simplest high-dimensional controlled operation as it only includes two-dimensional states as control and target states. However, as both, the radial and azimuthal degree of freedom are (in principle) unbounded, high-dimensional controlled operations can also be implemented using a single photon and qudit states encoded in both transverse degrees of freedom.

\section{Complete collection of measurements}
In order to show the flexibility of the presented technique, we performed a vast amount of measurements on different degrees of freedom, investigating different high dimensional transformations. The three tables below show all performed measurements on a three (table \ref{tab:1}) and a five dimensional (table \ref{tab:2}) Hilbert space of LG modes taking into account only the azimuthal index $l$ as well as on two- and three dimensional radial mode $p$-spaces (table \ref{tab:3}). The results shown below were obtained using eight phase-manipulation planes as well as generation- and detection holograms displayed on a single SLM. As a figure of merit we give the average visibility (eq.\ref{eq:Vis}) calculated from the correlation matrices. The most interesting and elucidating results are displayed in the figures and discussed in detail in the main text.

\subsection{Results 3D l-space transformations}

\begin{table}[htbp]
\begin{center}
\begin{tabular}{c||c|c|c}
3-D \\
\hline
\hline
 Gate   &$\hat{X^0}$    &$\hat{X^1}$  &$\hat{X^2}$  \\
 \hline
 Visibility &94.2~\% &96.5~\%  &87.5~\%  \\
\hline
\hline
 Gate   &$\hat{H_2}\hat{X^3}$  &$\hat{H_3}\hat{X^2}$  &$\hat{H_1}$  \\
 \hline
 Visibility  &96.2~\%   &92.9~\%   &85.6~\%  \\
\hline
\hline
Gate    &$\hat{H_1}\hat{X^1}$ & $\hat{H_1}\hat{X^2}$  &$\hat{H}_2^\dagger\hat{X}^{1(a)}$ \\
\hline
Visibility&86.7~\% &91.0~\%  &88.8~\%  \\
\hline
\hline
Gate     &$\hat{H}_3^\dagger\hat{X}^{1(b)}$  &$\hat{H}_1^\dagger\hat{X}^{1(c)}$  &$\hat{H}_2^\dagger\hat{X}^{1(d)}$  \\
\hline
Visibility  &92.9~\%  &95.9~\%  &91.9~\% 
\end{tabular}
\end{center}
\caption{All measured results for the transformations in a three dimensional $l$-only Hilbert space of LG modes ($p=0$). The gates marked with a superscript are different combined $\Hat{H}\Hat{X}$-gate transformations performed between different MUBs not starting in the computational basis (i.e. (a) MUB3 $\xrightarrow{}$ COMP (MUB1), (b) MUB4 $\xrightarrow{}$ COMP (MUB1), (c) MUB2 (ANG) $\xrightarrow{}$ COMP (MUB1) and (d) MUB4 $\xrightarrow{}$ MUB2 (ANG)). This further shows the flexibility and versatility of the presented technique.}
\label{tab:1}
\end{table}
\newpage
\subsection{Results 5D l-space transformations}
\begin{table}[htbp]
\begin{center}
\begin{tabular}{c||c|c|c}
5-D \\
\hline
\hline
Gate   &$\hat{X^0}$&$\hat{X^1}$  &$\hat{X^2}$   \\
 \hline
 Visibility &94.6~\%&	96.5~\%&	91.5~\%\\
\hline
\hline
 Gate   &$\hat{X^3}$  &$\hat{X^4}$  &$\hat{H_2}$  \\
 \hline
 Visibility &	87.9~\%&	89.2~\%& 76.5~\%
\end{tabular}
\end{center}
\caption{All measured results for the transformations in a five dimensional l-only Hilbert space of LG modes. Note that, as mentioned in the main text, the performance for quantum fourier transformations ($\hat{H}$-gates) decrease significantly due to the limited resolution of the holograms.}
\label{tab:2}
\end{table}

\subsection{Results 2D/3D p-space transformations}
\vspace{1cm}
\begin{table}[htbp]
\begin{center}
\begin{tabular}{c||c|c|c|c}
2-D \\
\hline
\hline
Gate$^1$ & $\hat{X^0}$ & $\hat{X^1}$  \\
\hline
Visibility$^1$ & 91.8~\% &	91.0~\% \\

3-D\\
\hline
\hline    
 Gate  &$\hat{X^0}$&$\hat{X^1}$  &$\hat{X^2}$    \\
 \hline
Visibility  & 90.8~\% & 96.2~\% & 98.4~\% \\
\hline
\hline
Gate  &$\hat{H_2}\hat{X^1}$  &$\hat{H_1}\hat{X^1}$ &$\hat{X}^{1(a)}$ &$\hat{H}_1^\dagger\hat{X}^{1(b)}$  \\
 \hline
Visibility  & 88.6~\% & 88.6~\% & 96.3~\% & 89.9~\%
\end{tabular}
\end{center}
\caption{All measured results for the transformations in two- and three dimensional p-only Hilbert spaces. Again, gates marked with a superscript  are transformations performed on superposition states (Input states are not prepared in the computational basis, namely (a) MUB3 $\xrightarrow{}$ MUB3 and (b) MUB2 (ANG) $\xrightarrow{}$ COMP (MUB1)).}
\label{tab:3}
\end{table}
\newpage

\end{appendix}

\bibliography{Bib}

\end{document}